\documentclass[12pt,floatfix,pre]{revtex4}
\usepackage{subfigure}
\usepackage{psfrag}
\usepackage{graphicx}
\usepackage{dcolumn}
\usepackage{bm}
\usepackage{natbib}
\usepackage{amsmath, amsthm, amssymb}
\def\strutdepth{\dp\strutbox}
\def\nw#1{\strut\vadjust{\kern-\strutdepth\vtop to0pt{\vss\hbox to\hsize
{\hskip\hsize\hskip5pt$\leftarrow$\hss\strut}}}{\em #1}}

\begin{document}

\title{
M-brane singularity formation
     }

\author{
J.Eggers$^1$ and J.Hoppe$^2$
}

\affiliation{
$^1$School of Mathematics, 
University of Bristol, University Walk,
Bristol BS8 1TW, United Kingdom  \\
$^2$Department of Mathematics, Royal Institute of Technology, 
100 44 Stockholm, Sweden
        }
\begin{abstract}
We derive self-similar string solutions in a graph representation,
near the point of singularity formation, which can be shown to 
extend to point-like singularities on M-branes, as well as to the radially
symmetric case. 
\end{abstract}

\maketitle
\section{Introduction}
For more than 40 years, \cite{BC66,BC67,BC68} (see also \cite{Witham}), 
various ways of solving the non-linear equation 
\begin{equation}
\ddot{z}(1+z'^2) - z''(1-\dot{z}^2) = 2\dot{z} z' \dot{z}'
\label{BI}
\end{equation}
are known. 
Recent work on higher dimensional time-like zero mean curvature
hypersurfaces includes \cite{Christo,Milbredt,Ho08,BNO08}
Here we show that (\ref{BI}) can develop singularities 
in finite time, starting from finite initial data. The structure 
of these singularities is described by the self-similar ansatz
\begin{equation}
z(t,x) = z_0 - \hat{t} + \hat{t}^{\alpha}h
\left(\frac{x}{\hat{t}^{\beta}}\right) + \dots
\label{selfs}
\end{equation}
where $\hat{t}:=t_0-t\rightarrow 0$ (the dots are indicating lower 
order terms). Inserting (\ref{selfs}) into (\ref{BI}) one finds
the above ansatz to be consistent provided 
$\beta=(1+\alpha)/2 > 1$, and
\begin{equation}
h''\left(2\alpha h - \frac{(\alpha+1)^2}{4}\xi^2\right) =
(\alpha-1)\left[h'^2 + \alpha h - 
\frac{3}{4}(\alpha+1)\xi h'\right]\; ,
\label{sim_equ}
\end{equation}
for (\ref{selfs}) to be an asymptotic solution of (\ref{BI}). 
For consistency with a finite outer solution of (\ref{BI}), 
the profile $h$ must satisfy
\begin{equation}
h(\xi)\propto A_{\pm}\xi^{\frac{2\alpha}{\alpha+1}} \quad 
\mbox{for} \quad \xi\rightarrow\pm\infty
\label{growth}
\end{equation}
(for a general discussion of matching self-similar solutions to the 
exterior see \cite{EF08}).

The ansatz (\ref{selfs}) is formally consistent for a continuum 
of similarity exponents $\alpha\ge 1$ and for any solution of the similarity
equation (\ref{sim_equ}). However, by considering the regularity of solutions 
of (\ref{sim_equ}) in the origin $\xi=0$  the similarity 
exponent must be one of the sequence 
\begin{equation}
\alpha = \alpha_n = \frac{n+1}{n}, \quad n\in \mathbb{N},
\label{exp}
\end{equation}
certainly if $h(0)=0=h'(0)$, and presumably in general 
(i.e. {\it all} relevant solutions of (\ref{sim_equ})). 
Of this infinite sequence, we believe that only $\alpha = 2$ is 
realized for generic initial data; indeed, in this case 
\begin{equation}
\xi = \zeta + c\zeta^3/3 , \quad h(\xi) = \zeta^2/2 + c\zeta^4/4,
\label{sim_sol}
\end{equation}
which we will deduce from a parametric string solution corresponding to 
(\ref{BI}). 

The importance of the similarity solution (\ref{selfs}) lies in the 
fact that it can be generalized to arbitrary dimensions, in particular
to membranes. We find that the same type of singular solution 
is observed in any dimension, even having the same spatial structure 
(\ref{sim_sol}).

\section{The similarity equation}
A way of satisfying (\ref{sim_equ}) is to demand
\begin{equation}
L^2:=h'^2 + 2\alpha h - (\alpha + 1)\xi h' = 0.
\label{sim_r}
\end{equation}
(differentiating e.g. $(1+\alpha)\xi = h' + 2\alpha h/h'$ 
one can eliminate $h''$, reducing (\ref{sim_equ}) to an identity,
as long as $h' \ne 1$). 

The transformation 
\begin{equation}
h(\xi) = \xi^2 g(\xi) = \xi^2\left(\frac{(1+\alpha)^2}{8\alpha}
-\frac{v^2}{2\alpha}\right)
\label{trans}
\end{equation}
yields
\begin{equation}
-\frac{d\xi}{\xi} = \frac{vdv}{v^2\pm\alpha v+ (\alpha^2-1)/4} = \frac{1}{2}
\left(\frac{\alpha+1}{v\mp\frac{\alpha+1}{2}} -
\frac{\alpha-1}{v\mp\frac{\alpha-1}{2}}\right)dv, 
\label{solve1}
\end{equation}
i.e. (choosing the upper sign)
\begin{equation}
\frac{\left|v-(\alpha+1)/2\right|^{\alpha+1}}
{\left|v-(\alpha-1)/2\right|^{\alpha-1}} = \frac{E}{\xi^2}.
\label{solve2}
\end{equation}

This yields solutions $v\in [(\alpha-1)/2,(\alpha+1)/2)$,
\begin{eqnarray}
\label{lim}
&&v\approx\frac{\alpha-1}{2} + \left(\frac{\xi^2}{E}\right)^
{\frac{1}{\alpha-1}} + \dots\quad \mbox{as}\quad \xi\rightarrow 0 \\ \nonumber
&&v\approx\frac{\alpha+1}{2} - \left(\frac{\xi^2}{E}\right)^
{\frac{1}{\alpha+1}} + \dots\quad \mbox{as}\quad \xi\rightarrow \pm\infty,
\end{eqnarray}
i.e. 
\begin{eqnarray}
\label{limits}
&&h(\xi)\ge 0, h(0)=0, \\ \nonumber
&&h(\xi)\propto \xi^2/2 \quad \mbox{as}\quad \xi\rightarrow 0 \\
&&h(\xi)\propto \frac{1+\alpha}{2\alpha}
   \xi^{\frac{2\alpha}{1+\alpha}} \nonumber \quad \mbox{as}\quad 
\xi\rightarrow \pm\infty. \nonumber
\end{eqnarray}
Note that these solutions are consistent with the growth conditions
(\ref{growth}). 

To solve the {\it second order equation} (\ref{sim_equ}) we note that 
$\tilde{h}(\xi):=c\; h(\xi/\sqrt{c})$ solves (\ref{sim_equ}), if $h$ does, 
and that 
\begin{equation}
\frac{h'}{\xi} - \frac{2h}{\xi^2} = \frac{1}{\alpha}
f\left(\sqrt{\frac{(\alpha+1)^2}{4}-2\alpha\frac{h(\xi)}{\xi^2}}\right) 
\quad \equiv \left(\frac{1}{\alpha}f(v)\right)
\label{reduce}
\end{equation}
reduces (\ref{sim_equ}) to 
\begin{equation}
-\left(v^2-\frac{(\alpha+1)^2}{4}\right)
\left(v^2-\frac{(\alpha-1)^2}{4}\right) = 
f\left(\alpha vf' - (\alpha-1)f-(\alpha+2)v^2+(\alpha^2-1)(\alpha-2)/4\right),
\label{reduce1}
\end{equation}
and 
\begin{equation}
\frac{d\xi}{\xi} = -\frac{vdv}{f(v)}=\frac{\alpha dg}{f}.
\label{reduce2}
\end{equation}

The growth condition (\ref{growth})
implies that $h$ grows less than quadratically at infinity. 
Thus we deduce from (\ref{reduce}) that $f$ vanishes at 
$(\alpha+1)/2$. Furthermore, from a direct calculation using the growth 
exponent (\ref{growth}) we find the first derivative, yielding
the initial conditions 
\begin{equation}
f\left(\frac{\alpha+1}{2}\right) = 0, \quad
f'\left(\frac{\alpha+1}{2}\right) = 1 .
\label{growth1}
\end{equation}
Using (\ref{growth1}), (\ref{reduce1}) yields a polynomial solution
\begin{equation}
f(v) = \left(v-\frac{(\alpha+1)}{2}\right)
\left(v-\frac{(\alpha-1)}{2}\right) = v^2-\alpha v + \frac{\alpha^2-1}{4},
\label{reduce3}
\end{equation}
(i.e.) (\ref{solve2}), which corresponds to the first order 
equation (\ref{sim_r}), but also an infinity of other solutions
(a Taylor expansion around $v_{\infty}=(\alpha+1)/2$ shows that 
(\ref{reduce1}) leaves $f''((\alpha+1)/2)$ undetermined, when 
(\ref{growth1}) holds). 
We note that (\ref{reduce1}) also has the solution $f_-(v) = f(-v)$, 
and for the special case $\alpha=2$ another pair of polynomial solutions, 
\begin{equation}
\tilde{f}(v) = \left(v+3/2\right)\left(v-1/2\right) = 
v^2 + v - 3/4,
\label{reduce4}
\end{equation}
and $\tilde{f}_-(v)=\tilde{f}(-v)$. While the asymptotic behavior 
following from (\ref{reduce4}) is in disagreement with (\ref{growth}), 
integration methods similar to those used by Abel \cite{Abel} 
perhaps permit a complete reduction of (\ref{reduce1})
to quadratures. 

In any case, (\ref{reduce1}) can be simplified in various ways. For
$\alpha=2$, e.g. it reduces to 
\begin{equation}
yy' = y-\frac{1}{4v^{5/2}}(v^2-9/4)(v^2-1/4)
\label{reduce5}
\end{equation}
via 
\begin{equation}
f(v) = \sqrt{v}y(4v^{3/2}/3).
\label{trans1}
\end{equation}

The solution (\ref{reduce3}), which is consistent with the growth 
condition (\ref{growth}), is equivalent to the solution (\ref{solve2}) 
of (\ref{sim_r}) given before. If one investigates the behavior of
the solution in the origin (either using (\ref{solve2}) directly or
by series expansion of (\ref{sim_r})), one finds that only for 
$\alpha = \alpha_n$ (cf. (\ref{exp})) a smooth solution is possible.
Thus the first consistent solution is found for $n=1$ or $\alpha = 2$. 
Higher order solutions $n = 2,3, \dots$ are also possible in 
principle. They have the property that apart from $f''(0)$, the first
non-vanishing derivative is $f^{(2n+2)}(0)$. However, 
we believe that they correspond to non-generic initial conditions,
whose derivatives have corresponding properties of vanishing up to a 
certain order. 
To demonstrate this point, one would have to perform a stability
analysis of the corresponding solution \cite{EF08}. In the string
picture discussed below this can be shown explicitly, as higher 
order solutions correspond to non-generic initial data. 

\section{Higher dimensions}
The solutions of (\ref{sim_r}), found to govern singularities of 
(\ref{BI}), also apply to higher dimensions. The reason is that 
the left hand side of (\ref{sim_r}) is the leading order term of
\begin{equation}
{\cal L}^2 = 1-\dot{z}^2+z'^2.
\label{lo}
\end{equation}
In other words, the asymptotic singular solutions discussed above
have ${\cal L}^2=0+$ lower order. In fact, differentiating (\ref{lo}) with 
respect to $t$ and $x$ one easily shows that ${\cal L}^2=0$ provides 
solutions of (\ref{BI}). In higher dimensions, differentiating 
$1 - z^{\alpha} z_{\alpha}=0$ gives $z^{\alpha}z_{\alpha\beta}=0$,
and hence 
\begin{equation}
(1-z_{\alpha}z^{\alpha})\square z + 
z^{\beta}z^{\alpha}z_{\alpha\beta} = 0. 
\label{membrane}
\end{equation}
Thus solutions of ${\cal L}^2=0$ also solve the M-brane equation
(\ref{membrane}) in arbitrary dimensions. 

For the special case of radially symmetric
membranes:
\begin{equation}
\ddot{z}(1+z'^2) - z''(1-\dot{z}^2) - 2\dot{z} z' \dot{z}' = 
\frac{z'}{r}\left(1-\dot{z}^2+z'^2\right) \equiv 
\frac{z'}{r}{\cal L}^2.
\label{BI_rad}
\end{equation}
Insert the radial version of the ansatz (\ref{selfs}), 
\begin{equation}
z(t,r) = -\hat{t} + \hat{t}^{\alpha}h
\left(\frac{r-r_0}{\hat{t}^{\beta}}\right) + \dots,
\label{selfs_rad}
\end{equation}
into (\ref{BI_rad}). If $r_0\ne 0$, the entire right hand side
of (\ref{BI_rad}) is of lower order in $\hat{t}$, and the similarity 
equation (\ref{sim_equ}) remains the same. Geometrically, this corresponds
to the singularity forming along a circular ridge of radius $r_0$. 

If on the other hand $r_0=0$, i.e. the singularity forms along the 
axis, the right hand side is of the same order, and the similarity 
equation becomes 
\begin{equation}
h''\left(2\alpha h - \frac{(\alpha+1)^2}{4}\xi^2\right) +
(1-\alpha)\left[h'^2 + \alpha h - \frac{3}{4}(\alpha+1)\xi h'\right] = 
-\frac{h'}{\xi}\left[h'^2 + 2\alpha h - (\alpha + 1)\xi h'\right].
\label{sim_equ_rad}
\end{equation}
This equation can in principle have solutions different from 
(\ref{sim_equ}). For solutions of (\ref{sim_r}), however, the 
expression in angular
brackets in (\ref{sim_equ_rad}) vanishes, hence solutions of 
(\ref{sim_r}) also solve (\ref{sim_equ_rad}). Thus 
(\ref{selfs_rad}),(\ref{sim_sol}) describe a point-like 
singularity on a membrane. 
These observations straightforwardly generalize to higher
M-branes, $M>2$.

\section{Parametric string solution}
\label{parametric}
Let us now compare our findings with the solution of closed 
bosonic string motions given by equation (50) of \cite{Ho95}.
(note that the definitions of $f$ and $g$ are changed by $\pi/4$, and
that the constant $\lambda$ is chosen to be 1):
\begin{equation}
\dot{{\bf x}}(t,\varphi) = \sin(f-g)\left(\begin{array}{c}
            -\sin(f+g)\\
            \cos(f+g)
            \end{array}\right)
\label{exact1}
\end{equation}
\begin{equation}
{\bf x}'(t,\varphi) = \cos(f-g)\left(\begin{array}{c}
            \cos(f+g)\\
            \sin(f+g)
            \end{array}\right),
\label{exact2}
\end{equation}
where $f = f(\varphi + t)$ and $g = g(\varphi - t)$. From 
(\ref{exact2}) one finds the curvature 
\begin{equation}
k(t,\varphi) = \frac{f'+g'}{\cos(f-g)} .
\label{curv}
\end{equation}
The hodograph transformation 
\begin{eqnarray}
\label{hodo}
&& (t,\varphi)\rightarrow t=x^0, \quad x = x^1(t,\varphi), \\
&& x^2(t,\varphi) = z(t,x^1(t,\varphi)), \nonumber
\end{eqnarray}
implying $\dot{z}=\dot{y}-\dot{x}y'/x'$, $z'=y'/x'$,
$(\partial\phi/\partial x^0=-\dot{x}/x', 
(\partial\phi/\partial x^1=1/x')$
permits to go between the parametric string picture 
(\ref{exact1})-(\ref{curv}) and the graph description (\ref{BI}). 
In particular, 
\begin{equation}
1 - \dot{z}^2+z'^2 = \left(\frac{\cos(f-g)}{\cos(f+g)}\right)^2
\label{cone}
\end{equation}
is manifestly non-negative in the parametric string-description, 
while for solutions of (\ref{BI}) one has to demand it explicitly -
leading e.g. to the exclusion of solutions with $h'(0)=0$, 
$h(0)<0$. 

Let us give an explicit example of 
$\mathbb{M}_2 \subset \mathbb{R}^{1,2}$ being at $t = 0$ a 
regular graph, while for $t= 1$ a curvature singularity 
has developed. Let $\mathbb{M}_2$ be described by ${\bf x}(\varphi,t)$, 
as defined by (\ref{exact1}),(\ref{exact2}), with $\varphi\in\mathbb{R}$,
$t\ge 0$. Let 
\begin{eqnarray}
\label{example}
f(w)=\left\{\begin{array}{ll}
             \arctan w \quad &\mbox{for} \quad w\ge\epsilon \\
              \chi_{\epsilon}(w)\arctan w 
                &\mbox{for} \quad 0\le w\le \epsilon < 0 \\
              0 &\mbox{for} \quad w\le 0 
                 \end{array}\right.,
\end{eqnarray}
where $\chi_{\epsilon}(w\ge \epsilon)=1$, $\chi_{\epsilon}(w\le 0)=0$, 
and $\chi_{\epsilon}(0<w< 0)$ such that $f'(w)\ge 0$.
We also assume that $g(w)=-f(-w)$. 
A simple calculation then shows that for 
$\varphi\in [-t+\epsilon,t-\epsilon]$ one obtains 
(${\bf x}_0(t=0,u=0)=0$) 
\begin{eqnarray}
\label{explicit}
&& x(\varphi,t) = -\varphi+\arctan(\varphi+t)+\arctan(\varphi-t)  \\ \nonumber
&& y(\varphi,t) = \ln\sqrt{\left(1+(\varphi+t)^2\right)
\left(1+(\varphi-t)^2\right)} \\
&& k(\varphi,t) = \frac{2}{\sqrt{\left(1+(\varphi+t)^2\right)
\left(1+(\varphi-t)^2\right)}}\frac{\varphi^2+1+t^2}{\varphi^2+1-t^2} \nonumber
\end{eqnarray}
Note that for $t>1$ this is no longer a graph. 

\begin{figure}
\includegraphics[width=0.5\hsize]{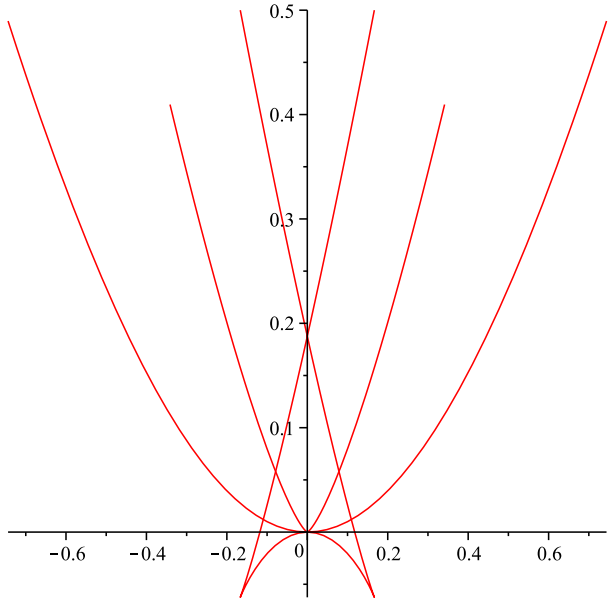}
\caption{
The formation of a swallowtail, as described by (\ref{BI_swallow}). Shown is 
a smooth minimum, ($\hat{t} > 0$), a minimum with a 4/3 singularity
($\hat{t} = 0$), and a swallowtail or double cusp ($\hat{t} < 0$). 
   }
\label{swallow_fig} 
\end{figure}
The example (\ref{example}) underlies a general structure that 
can be uncovered by a local expansion of the functions $f$ and $g$
around the singularity. Namely, as seen from (\ref{curv}), the singularity
occurs when $f-g$ is a multiple of $\pi/2$; for simplicity, we also 
assume that $g(w)=-f(-w)$ as before. 

Then a local Taylor expansion yields
\begin{equation}
f(\zeta) = \pi/4 + a(\zeta-\zeta_0) - b(\zeta-\zeta_0)^2 + 
O(\zeta-\zeta_0)^3,
\label{f_exp}
\end{equation}
so together with the symmetry requirement we find 
\begin{equation}
f - g = \pi/2 + 2a(t-\zeta_0) - 2b\varphi^2 + (t-\zeta_0)^2, 
\label{f_minus}
\end{equation}
where we neglected higher-order terms in the expansion, which 
will turn out to be irrelevant for the singularity formation.

From this expression it is clear that $\zeta_0$ has to be identified 
with the singular time $t_0$ and $a>0$ for the solution to be regular 
for $t<t_0$. 
Similarly, one must have $b>0$ (otherwise $f-g$ would be $\pi/2$ at an 
earlier time), and the singularity occurs at $\varphi = 0$. Thus to 
leading order we have 
\begin{equation}
f - g \approx \pi/2 - 2a \hat{t} - 2b\varphi^2, \quad 
f + g \approx 2a\varphi, 
\label{f_sum}
\end{equation}
from which we get 
\begin{equation}
x' = 2a\hat{t} + 2b\varphi^2, \quad y' = 4a^2\hat{t}\varphi + 4ab\varphi^3.
\label{xp}
\end{equation}
Integrating this expression, using the integrability condition
(\ref{exact2}), gives 
\begin{equation}
x = \hat{t}\varphi + c\varphi^3/3, \quad y = -\hat{t} + 
\hat{t}\varphi^2/2 + c\varphi^4/4,
\label{BI_swallow}
\end{equation}
where we have used a rescaling of the parameter $\varphi$. 

The curve described by (\ref{BI_swallow}) is shown in Fig.~\ref{swallow_fig},
but disregarding the spatial translation of $z$ by the term $-\hat{t}$.
In catastrophe theory, this is known as the swallowtail \cite{Arnold84}. 
For $\hat{t} > 0$ the curve is smooth, while for $\hat{t} = 0$ a rather
mild singularity develops; at the origin, $y\propto x^{4/3}$. 
After the singularity ($\hat{t}<0$) the curve self-intersects. The solution
(\ref{BI_swallow}) leads directly to the similarity form (\ref{selfs}),
if one notices that $\varphi$ is of order $\hat{t}^{1/2}$ for the terms in 
(\ref{BI_swallow}) to balance. Thus, using the notation of (\ref{selfs}),
and putting $\zeta=\varphi/\hat{t}^{1/2}$, one finds $\alpha = 2,
\beta = 3/2$ for the exponents, and (\ref{sim_sol}) 
for the similarity function. The crucial point is that although 
(\ref{sim_sol}) came out of an expansion, all higher order terms
are subdominant in the limit $\hat{t}\rightarrow 0$. Thus 
(\ref{sim_sol}) is in fact an {\it exact} solution of (\ref{sim_equ})
with $\alpha = 2$. Moreover, it is even a solution of (\ref{sim_r}), 
and precisely of the form (\ref{solve2}). 

In the case of non-generic initial conditions other solutions are
possible. For example, instead of (\ref{f_exp}) 
\begin{equation}
f(\zeta) = \pi/4 + a(\zeta-\zeta_0) - b(\zeta-\zeta_0)^{2n},
\label{f_exp_gen}
\end{equation}
where $n\in\mathbb{N}$ but $n>1$. Only even powers $2n$ are allowed, 
otherwise the singularity occurs for all $\varphi$ at the same time,
i.e. it is no longer point-like. If the leading order term is not
linear but itself of higher order, the resulting similarity profile 
becomes singular at the origin, cf. \ref{sim_sol_high}. 
Repeating the above calculation 
along the same lines, we find 
\begin{equation}
x' = 2a\hat{t} + 2b\varphi^{2n}, \quad 
y' = 4a^2\hat{t}\varphi + 4ab\varphi^{2n+1},
\label{xp_non}
\end{equation}
which is equivalent to the symmetric shape function 
\begin{eqnarray}
\label{sim_sol_non}
&& \xi = \zeta + 2d(n+1)\zeta^{2n+1}/(2n+1)  \\ \nonumber
&& h = \zeta^2/2 + d\zeta^{2n+2}.
\end{eqnarray}
The corresponding similarity exponent is $\alpha=\alpha_n$, as given
by (\ref{exp}). We thus retrieve the exact same solutions identified
by our previous analysis, based on a similarity description. 

\appendix*
\section{Additional solutions of the similarity equation}
It is possible to construct many more solutions to the similarity
equation (\ref{sim_r}), which are all defined on the real line, 
but which we reject since they either contradict (\ref{growth})
or are not smooth. The simplest case is 
\begin{equation}
h(\xi) = \frac{\xi^2}{2},
\label{sol_simple}
\end{equation}
which is a solution for {\it any} $\alpha$, but evidently does
not satisfy the matching condition (\ref{growth}). 

Recall that for $\alpha=\alpha_n$, (\ref{solve2}) furnishes smooth 
solutions on the real line. On the other hand, while for $\alpha=3$ the second
derivative of the resulting solution is well-defined, the third 
derivative is discontinuous. Namely, for $E=4$ (e.g.) one finds that 
for $\xi>0$, 
\begin{eqnarray}
\label{special3}
&&h(\xi) = -\frac{2}{3}\left(\xi+2(1+\xi)-2(1+\xi)^{3/2}\right) \\ \nonumber
&&h'(\xi) = 2\left(\sqrt{1+\xi}-1\right) > 0 \\ 
&&h''(\xi) = 1/\sqrt{1+\xi} \nonumber,
\end{eqnarray}
so that 
\begin{eqnarray}
\label{case3}
h'''(\xi)=\left\{\begin{array}{ll}
             -(1+\xi)^{-3/2}/2 \quad &\mbox{for} \quad \xi > 0\\
             (1+\xi)^{-3/2}/2 \quad &\mbox{for} \quad \xi < 0. \\
                 \end{array}\right.
\end{eqnarray}

Other solutions whose scaling exponent is not from the set (\ref{exp}),
but which have well-defined second derivatives, can be found from the 
parametric string solution as described in section~\ref{parametric}. 
If the expansion of $f$ does not start with a linear term as in 
(\ref{f_exp}), but at higher order, e.g. 
\begin{equation}
f(\zeta) = \pi/4 + (\zeta-\zeta_0)^3/2 - b(\zeta-\zeta_0)^4,
\label{f_higher}
\end{equation}
one finds instead of (\ref{xp})
\begin{equation}
x' = 3\hat{t}\varphi^2 + 2b\varphi^4, \quad 
y' = 3\hat{t}\varphi^5 + 2b\varphi^7.
\label{xp_high}
\end{equation}

Integrating (\ref{xp_high}), the result once more conforms with
(\ref{selfs}), with a similarity exponent of $\alpha =4$,
and the similarity function has the parametric form
\begin{eqnarray}
\label{sim_sol_high}
&& \xi = \zeta^3 + 2b\zeta^5/5  \\ \nonumber
&& h = \zeta^6/2 + b\zeta^8/4.
\end{eqnarray}
It is confirmed easily that (\ref{sim_sol_high}) solves (\ref{sim_r}) 
with $\alpha=4$, but the third derivative of $h(\xi)$ is singular
at the origin. 

\begin{acknowledgments}
J.H. would like to thank P.T. Allen and J. Isenberg
for a discussion and correspondence, as well as
the Swedish Research Council and the Marie Curie
Training Network ENIGMA (contract MRNT-CT-2004-5652)
for financial support.

\end{acknowledgments}

\end{document}